# Creating a universe from thermodynamic thought experiments II - exploring a basis for structure and force.

Akinbo Ojo

**Abstract**

Assuming that the universe is a system obedient to known thermodynamic laws and equations, here we explore whether it is a possibility for the universe to exist and evolve without any cosmic structure or force necessarily emerging within it. From symmetry considerations and the invariance of Boltzmann's constant during the experiment, we infer that it is an inevitable occurrence that structure and force will appear during transformation of the created universe.

**Introduction**

In an earlier experiment [1], we created an increase in the size of an infinitesimal isolated system using thermodynamic tools derived from Clausius and Boltzmann's descriptions of entropy [2-6]. We called this system our universe and now explore whether as a theoretical speculation, it is at all a possibility for a universe to exist and increase in size without any structure or force necessarily evolving within it. Our main objective will be to find a qualitative rather than a quantitative basis for the existence of force and structure. The experiment will again make use of Clausius's description of entropy expressed by

$$\Delta E/T = \Delta S \qquad (1)$$

and that of Boltzmann expressed by

$$S = k \ln W \qquad (2)$$

where S = entropy, E = energy, T = absolute temperature, *k* = Boltzmann's constant, ln stands for natural logarithm (i.e. $\log_e$, where e = 2.71828…) and *W* represents the number of possible ways the constituents can be arranged among the compartment units of the system.

To make the results of our experiment clear, we first do a brief but pertinent description of the objects of our search, force and structure, lest we see but miss them. In embarking on this search we also take a cursory look at what *W* means for our experimental endeavor.

Even though *W*, structure and force can be described in various ways, the descriptions we employ will be biased towards the thermodynamic perspective of the area of interest. Let us commence with an illustrative description of *W*.

Consider a system of three compartments with three different constituents, say three boxes and three balls, Fig.1. Each compartment unit is made as small as to be able to harbor only one constituent and no more.

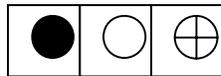

Fig.1. A system made up of three compartments and three constituents.

In this sample system, if we exclude the mirror images for simplicity, the system can be arranged in three possible ways as shown below, i.e. *W* = 3 and entropy S is proportional to the natural logarithm of that number according to Eq.(2).

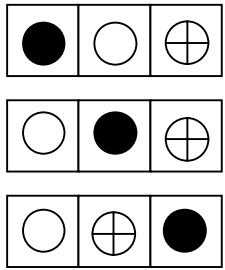

Fig.2. Illustrating *W*, the number of possible ways the system can be arranged.

Note that in this example, moving the balls for arrangement gives the same number of possible ways of arranging the system as moving the boxes with their contained balls and so in calculating the value of $W$ it makes no difference whether we regard the boxes or the balls as the constituents or subjects of arrangement. If the boxes are so small that they cannot be further divided to contain more than one constituent, then for all the practical purposes of calculating $W$ the boxes can be taken to serve both as compartment units and as the constituents that can be arranged in the various possible ways. Bearing this in mind we sometimes interchangeably use the terms constituent and compartment unit where this will help clarity of expression.

From the above, if it becomes mathematically compelling that we have to reduce $W$ in the system, then a constituent compartment or more will have to be excluded from further participating in the different possible arrangements in the system leaving available other constituent compartments before this reduction can be physically realized.

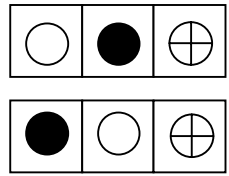

Fig.3. Illustrating what would have to occur if the number of ways of arranging the system in Fig.1 has to be reduced from three to two. A constituent or compartment unit, e.g. the one with the crossed ball will have to be excluded from further involvement in the different arrangements possible in the system in order for this reduction in $W$ to be possible.

It is easy to visualize from this diagrammatic representation that the more the number of compartment units and constituents available for arrangement, the higher $W$ will be in the system. Therefore, if we must have an increase in $W$, we either break up the constituents and compartments so that more constituents and compartment units become available and give us more ways of arranging the system, or if this is not feasible and yet it is thermodynamically compelling that $W$ has to be increased, then the system must come up with more compartment units to add to the 3-unit system we have, so that such mathematically demanded increase in the number of possible ways of arranging the system is physically realized.

When within a system in which various arrangement possibilities can take place, a particular group of constituents become excluded from these arrangement possibilities, we may envisage the emergence of the phenomenon of structure. Observationally, such phenomenon will be contrary to the second law of thermodynamics which stipulates that constituents of a system should not to be restricted but should rather increasingly participate in all and more of the different ways the system can be arranged so that we have a continuously increasing $W$. For analogy, we consider a glass of water of given volume, Fig.4. At an initial earlier time, the water molecules in Glass A are all fully participating and involved in the possible arrangements available in the system. From the volume of water and the number of molecules, we can even estimate a finite, but astronomical value for the number of ways the system can be arranged without affecting our description of the glass of water as a macroscopic state. If we extract some energy from this glass of water, then even though the number of molecules contained therein is unchanged, Eq.(1) tells us that we have reduced the entropy and $W$ must reduce. We see that if the amount of water remains the same, the only way a reduction in $W$ can be effected is to exclude some of the water molecules from further participating in all the possible arrangements in which they could hitherto partake. Such water molecules that are so excluded appear as structural forms and may be visualized as ice-crystals in Glass B.

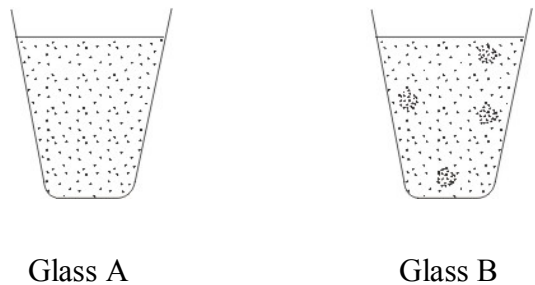

Glass A          Glass B

Fig.4. When energy is extracted from glass A, $W$ must reduce and some water molecules have to be restricted from further involvement in the different possible arrangements available to the system to make this possible. Such molecules no longer available for arrangement in all the possible ways can be visualized as ice crystals in glass B.

The above scenario is reversible so that if we add energy to the glass of water with the ice-crystals, $W$ has to increase in obedience to Eq.(1). To make this happen we need either more compartments in which arrangements can take place or more constituents that can be available to give more possible ways of arrangement, thus the water molecules contained in the ice structure must be yielded up for participation to give the additional possible arrangements that must become available in the system so that a higher value of $W$ is realized. More and more energy can be added till even the molecules may need to be broken up into even more fundamental constituents to satisfy any further demand for increases in $W$. With this brief analogy to illustrate a relationship between $W$ and the emergence of structure, we next examine the concept of force.

What is force? We seem to know what it does but we do not fully know what it is and why it must do what it does. What does a universe need it for? At the beginning when we were experimentally creating the universe [1], there was no force, so who could have asked for it since the presence of force is known to exist in the real experiment? Is it inevitable for it to emerge and why? The subject of force, a major preoccupation of Newton [7], would require a book of its own being a subject that can be viewed from many different and intriguing perspectives.

We know for instance that a system cannot experience any force unless it exchanges momentum with another system according to Newton's laws and this exchange must be both equal and in opposite directions. A bullet cannot exit a gun, unless it exchanges momentum with the gun, similarly a car cannot move unless it exchanges momentum with the road. Force thus seems non-existent if a system has no other system to relate with and a body with inertia-mass continues in its state of rest or uniform motion in a straight line* unless it exchanges momentum with another body and causes that body's deceleration before it itself can simultaneously accelerate and experience force. However this experience of momentum change or acceleration is according to Einstein's equivalence principle not very reliable to experimentally discern the presence of force. This is because according to the principle, acceleration depends on the observer's frame of reference. That is, if you set out to detect the existence of force through acceleration whether you find it or not depends on the frame of reference from which you are making your observation. This relative nature of observed force is well exemplified by the weightlessness experienced by astronauts in space. Galileo too famously demonstrated the difficulty of using acceleration to know the magnitude of force by showing that falling objects subjected to different amount of force can still experience the same acceleration. The moral is that acceleration is not totally reliable to experimentally discern the absence or presence of force. We need more reliable markers, which may not depend on an observer's frame of reference.

---

*As a corollary to this postulate of Newton, the state of rest or 'natural' motion of a massless body would then be random rather than uniform motion in a straight line since such a body would not need force to change its direction from a straight line, such changes in direction being forceless.

As a preliminary evaluation based on what we know of the universe in the real experiment, an atom's electron should be free to roam and be available for arrangement in the different possible ways like other particles present in a compartment, in obedience to the second law of thermodynamics. However if the electron refuses to obey this law and instead of being arranged in all the possible ways within the space of the compartment like other particles, it is seen to cling to another particle, force labeled electromagnetic is inferred to be present. Similarly, a planet is composed of particles which we would expect to scatter and be spread out randomly to occupy more and more space as $W$ continually increases as dictated by the second law. If the planet's particles refuse to obey this law and appear to be constrained from participating in all the different available arrangements of constituents in which they should otherwise be taking part, force labeled gravitational is used to explain the disobedience. Using our gun and bullet for analogy, we would expect the molecules of the bullet to intermingle, spread and mix among the molecules of the gun and if there is space, the molecules of the gun and bullet should further spread out into the room where they are kept in obedience to the second law. If however we do not see this, intermolecular forces are used to explain the absence of mixing between the molecules of the bullet and the gun while frictional force is used to explain why the bullet does not separate and slip out of the gun on its own to roam about the room. In Glass B of Fig.4 as another example, the presence of electrostatically based intermolecular forces called hydrogen bonds are used to explain the stability of ice and how the restriction of its contained water molecules from participation in the available arrangement possibilities in the glass is made a physical reality. Learning from the real universe, force should therefore be looked out for in situations where we encounter disobediencies of the second law of thermodynamics. In such situations $W$ would either be static or reducing and this would manifest either by a complete refusal to be arranged in more of the compartment volume or spreading out in an ordered, non-probabilistic and predictable way such that cannot be said to constitute an increase in *W* or disorder.

Before proceeding with our experiment we examine another signal that can help us decipher the presence of force. This is from Gauge theory which is a theory based on symmetry and the idea that symmetrical transformations can be performed locally and globally. Symmetry is said to exist when a measurable quantity remains invariant even when the system is undergoing change and transformation. Exploitation of symmetry has been rewarding for physicists in explaining natural phenomena. Historically, the energy conservation laws in thermodynamics, Einstein's invariance or relativity theory, Newton's momentum conservation in dynamics and Yang-Mills theory in particle and quantum physics have all found symmetry a fertile ground for discovering and explaining physical theories and concepts. Physicists have learnt to identify mathematically defined symmetrical relationships and to ask how these relationships on a global scale relate to the same relationship on a local scale during transformation. In other words, how does a global observance of symmetry, i.e. obedience to a symmetrical principle every time and everywhere relate with the local observance of the same symmetrical principle, i.e. obedience of symmetry at a given place and time? How does the local system "know" whether and in what way the global system is obeying the symmetry principle? This line of thought is probably original to Yang and the answer proposed by him was that the medium that communicates between local and global observance of an invariant principle was "force". Force is described as nature's way of expressing the global symmetry in local situations [8,9]. The global system need only observe the global symmetry but the local system is subject to both the local and global observance of symmetry. Where the symmetry is observed in the same way locally and globally there may possibly be no need for force to exist. However where during transformation the local and global systems are observing the same symmetrical principle in opposing ways, force would manifest locally as a witness to the conflict. Put in other words, whenever force is seen locally it is a sign that the global system is obeying a symmetrical principle in a way different from how the local system is obeying it.

As a familiar example, using the gun and bullet system again for analogy, we can identify conservation of momentum as the symmetrical principle which remains invariant during transformation and is obeyed globally and locally. Globally for a system at rest relative to the observer, momentum can be zero in the observational frame of reference. If during a transformation, a local part of the system, e.g. the gun, obeys the symmetrical principle in a way different from the global zero momentum way and instead has a local non-zero momentum in a particular direction, e.g. the direction of recoil, force would manifest as a witness to the conflict. Although this manifestation occurs as well in the gun, we commonly describe the manifestation of force in the bullet, which moves in the direction of fire with such velocity as to keep the symmetrical principle invariant.

In summary, the hallmarks we should be on the look out for as a basis for detecting the existence or emergence of force would seem to include scenarios where we encounter disobedience to the second law of thermodynamics and whenever there is disharmony in the way two interacting systems are obeying an overriding symmetrical principle which nevertheless remains locally and globally invariant no matter how the systems are transforming. Since structure and force seem associated with reduction or restraint of increase in $W$, in opposition to the dictates of the second law of thermodynamics we may expect from the foregoing that some form of co-existence and intimate relationship could exist between the two.

**The Experiment**

**Aims and Objectives:** (i) To investigate whether the emergence of force and structure is necessary or inevitable during the thermodynamic transformation of a created universe.
(ii) To investigate whether during the thermodynamic transformation of a created universe it is inevitable for some constituent compartments of the system to become excluded from participating in the different arrangements available to the system and in which they would otherwise have participated, necessitating the emergence of structural forms.
(iii) To investigate whether during the thermodynamic evolution of the universe, a disharmony exists in the way an invariant symmetrical principle is being obeyed by the global and local system, necessitating a need for force to emerge.

**Apparatus:** We employ a device capable of monitoring and measuring local and global changes in $W$ and other thermodynamic parameters based on Clausius and Boltzmann's descriptions of entropy as expressed by Eqs. (1) and (2), both of which give

$$\Delta E/T = \Delta S = k \, \Delta \ln W \qquad (3)$$

From this, we get

$$\Delta E/T\Delta \ln W = k \qquad (4)$$

which tells us more clearly how *W* will be expected to change in response to changes in E in any transformation. This invariance in Boltzmann's constant notwithstanding how a system is changing hints very strongly of symmetry. A positive change in E gives an increase in *W,* the number of ways the constituents of that system can be arranged, while a negative change and reduction in energy results in a reduction of the number of ways the constituents of the system can be arranged. The lower the temperature, T is at the initial time, the greater the impact any changes in E will have on *W*. From Eq.(4), no matter how E and *W* change, their mathematical relationship and Boltzmann's constant is invariant.

**Method:** We create a universe experimentally [1] by wishing a fluctuation in the energy of a closed system of zero or unit compartment and at an appropriately low temperature thereby forcing a significant change in its entropy and an astronomical change in *W*. As can be recollected *W* increases either if there is an increase in the number of constituents available for arrangement or if there is an increase in the number of compartment units among which arrangement can take place. In the earlier paper, methods of wishing a change in the energy of the closed system under experiment were suggested and include the uncertainty principle and quantum tunneling.

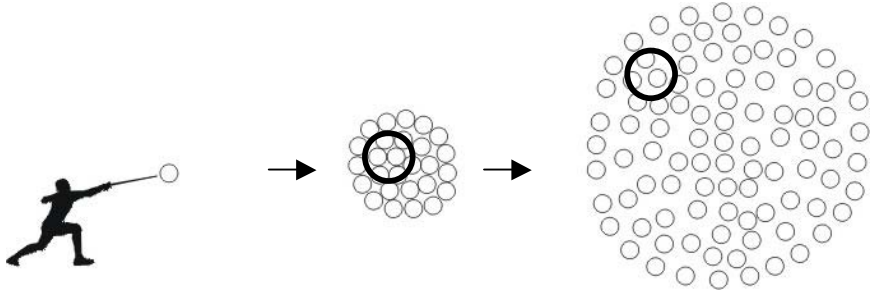

Fig.5. Illustrating what could occur if at a very low or zero temperature a phenomenon or an intelligent Agent were to cause an energy fluctuation, ΔE in a closed system infinitely so small it cannot be further fragmented into more compartments and yet *W* has to increase in obedience to Eqs. (1) and (3). The thermodynamic behavior of a sample local system (inset, in circle) is monitored during the occurring transformation of the global system.

Having set up our system, we use our device to monitor a particular sample local system of given volume, as the system evolves.

**Observation and experimental results:**

1. When energy E is introduced into the system, since the initial compartment unit cannot be further fragmented, new compartment units must physically appear to fulfill the mathematical demand for $W$ to increase and to avoid violations to Clausius' equation, Eq.(1). We observe an increase in the $W$ of the system caused by an initial change in E and this is manifest by a global proliferation of compartment units among which the demanded additional possible arrangements can take place.

2. The introduced energy E is present in the system as a whole but when the system increases in size, the amount of energy in our sample local system of given volume gets progressively less. If Eqs.(1) and (3) are to be preserved locally, then $W$ has to correspondingly reduce in the sample local system.

3. The reduction of $W$ in the sample local volume with time as the system evolves can only be manifest if some of the constituents that make up that volume are excluded from further participating in all the different available arrangements possible in the system. This satisfies the requirement for emergence of structure, as structure must originate in a scenario where some constituent compartments can be arranged in all the different ways possible while some are excluded and thermodynamically restrained from similar arrangement in all the available ways.

4. As $W$ reduces further locally in response to the continuing local reductions in energy as the universe expands, the number of constituents in that volume that are excluded from being arranged in all the available ways is found to be increasing with time.

5. From the thermodynamic scenario observed, it is found that a symmetrical relationship, Eq.(4), which is invariant during transformation is obeyed in opposing ways by the interacting local and global systems. During transformation there is a global tendency for $W$ to increase while obeying the symmetrical relationship described and a local tendency for $W$ to reduce while obeying the same symmetrical principle. The local reduction in $W$ that is observed is discordant with the global increase of $W$ dictated by Eqs. (1) and (3), which we also know as the second law of thermodynamics. The requirements for local display of force are thus satisfied according to the earlier identified criteria that would signal its emergence.

**Discussion:**

As the system is transformed and the need for the first constituents to be constrained is mandated by local reduction in *W*, the initial resulting structures in our thought experiment will be expected to have small constitutional arrangements since the local reduction in energy and *W* that necessitates their formation is still relatively small. Such structures being constituted during comparatively higher local energy regimes can be expected to be more stable to break up by energetic processes than latterly occurring ones, in the conventional parlance they can be said to have higher binding energies. We may therefore expect hierarchical manifestations of structure as the system is transformed, the smallest and most stable emerging before the larger and less stable to break up.

In the real experiment, the system also appears to have evolved from a beginning of very high energy density, the energy density reducing as the universe expanded subsequently permitting the emergence of quarks, leptons, hadrons, atomic, molecular and planetary structure in that order with characteristics of size and stability similar to the structures evolving in our thought experiment.

In our thought experiment, change in E is the cause of the change in *W* and *W* increases till the expected corresponding value, called the thermal equilibrium value is attained. Since this equilibrium value of *W* may not be attained instantaneously and yet the value of *k* must remain constant as given by Eq.(4), the value of T must vary from the initial to maintain that constant value of *k* during the transformation. Thus T which was very low or of zero value at the beginning of the experiment must become temporarily high. As *W* increases to its equilibrium value, T can then gradually return to its original value.

In the real experiment, even in models of a universe starting from nothing, an astronomically high temperature is described at or soon after the beginning. The temperature is taken to have been dropping with time, the system now having cooled to 2.7K, the current temperature of the cosmic microwave background radiation. That *W* has been increasing with time in the real universe is also a reality described by the second law. As the universe expands, the continuing reduction in local energy and temperature towards zero will as given by the third law of thermodynamics, i.e. S→ 0, when T→ 0, lead to reduction in local entropy, a scenario at variance with the global increase of entropy. We find therefore that in the real universe, the thermodynamic scenario locally differs from the scenario globally and we may be entitled to ask as Yang would do [8,9], how the global behavior is communicated to the local system?

We earlier identified hallmarks for the emergence of force. These are the presence of a symmetrical principle which remains invariant during transformation; the obedience of the invariant principle in opposing ways by the global and local system and the occurrence of possible violations to the second law of thermodynamics. If these conditions are found to be present, the phenomenon described as force should be reasonably expected. Though all forces seem to share these hallmarks, for the dynamical forces the invariant principle appears to be momentum conservation. On the other hand the fundamental forces in the real experiment, viz. gravitational, electromagnetic and the strong force all seem to be intimately connected with the maintenance of structure in disobedience to the dictates of the second law that constituents should be spread and arranged in increasingly more possible ways. This seeming association with the second law is suggestive that the applicable symmetrical principle could be thermodynamic in nature. From the results of our thought experiment, the invariant symmetrical principle which seems applicable for the emergence of force is the invariance of Boltzmann's constant and appears to differ from momentum conservation which applies to the dynamical forces.

As energy further reduces locally with accompanying reduction in $W$, there will be a need for further constraint to the possible arrangements of constituents available in continuing defiance of the second law but in obedience to the symmetrical principle. To make this mathematically demanded reduction in $W$ a physical reality, the system must utilize either an existing property or possibly induce the emergence of new properties in the units or group of units to be restrained from participation in the various arrangement possibilities, so as to distinguish them from the unconstrained constituents. For instance in Fig. 4, even though all water molecules contain hydrogen, the hydrogen in the ones to be constrained are endowed with "sticky" new properties and they make use of this to effect the realization of the mathematically demanded reduction in $W$. This they do by forming hydrogen bonds with other water molecules destined for confinement as ice.

For liquids that do not contain hydrogen, the system must device other physical means of effecting the constraint in $W$ when demanded. Such other devices apart from hydrogen bonds may still be electrostatic in nature, e.g. the ionic bonds that exist in polar molecules. For non-polar molecules, if it becomes thermodynamically imperative that some molecules have to be constrained, polarity is known to be inducible in hitherto non-polar molecules to give rise to other types of electrostatically based restraints, such as Van der Waal's forces, covalent and metallic bonds. It would therefore seem that properties possessed by constituents that distinguish and enable them to be constrained so that various structures can be formed must have evolved and have been induced as thermodynamic necessities. In our thought experiment, properties that describe the stronger constraints would evolve before weaker ones because they are necessitated in earlier higher energy regimes. The strength of a fundamental force based on those properties can then be measurable by how much energy would be required before the constrained constituents can be liberated to participate in the available arrangement possibilities. Also the need for properties that describe constraints of increasingly longer reach will similarly evolve as energy reduces even further locally with accompanying larger reduction in $W$ and increase in the number of constituent compartments that need to be restrained in the necessitated larger structures. We therefore expect that in our thought experiment, the initially evolving fundamental forces would be effected by properties that make them stronger and shorter in reach than latterly evolving ones. An important feature of the thought created universe would therefore be that such properties that give rise and are used to describe force are not fundamental or conserved but can similarly be lost or cancelled out if it becomes thermodynamically necessary.

In the real experiment the necessity for properties such as "color", charge and mass to evolve or be induced in those constituents to be constrained from participation in the various available arrangements in order that structure can be formed with the correspondingly described strong, electromagnetic and gravitational forces respectively would likely have become necessitated and evolved in similar order as in our thought experiment. The strong force being stronger and having less reach is for example likely to be described earlier than the weaker and longer reach forces. It is clear from what has been discussed that even if at any point in transformation, properties such as mass for gravitational force or charge for electromagnetic force become unable or insufficient to provide enough constraint and yet it becomes thermodynamically imperative for the system to further constrain constituents so that $W$ is even further reduced locally as demanded by the invariant relation in Eq.(4), some other means or type of force to describe that mathematical necessity must evolve. Theoretically there is nothing stopping further manifestations of such restraints and fundamental forces at different times as the system continues to expand with further reductions in local energy and even further need to reduce $W$ locally. We risk the speculation that such a force associated with galactic structures can evolve like others, even though galactic structure and stability is currently explained gravitationally with missing mass and dark matter [10].

The weakness of acceleration-deceleration as a reliable marker for detecting force has been mentioned in the introduction. Acknowledging this we can with care apply the concept to our description of the fundamental forces. In doing this, we take some liberty to broaden the definition of acceleration and identify it with the local or global direction in which the symmetrical principle is being obeyed. Using the gun and bullet system again for analogy, we can identify acceleration of the bullet with one of the directions in which the symmetrical principle is obeyed locally, i.e. the direction of fire, even though it is simultaneously decelerating to the opposite direction in which symmetry is locally obeyed, i.e. the direction of recoil. While the direction in which the global system is obeying the symmetrical principle can be identified with the direction of rest relative to the observer, i.e. zero acceleration. From the thermodynamic perspective and the invariant symmetrical principle found applicable to the fundamental forces in our thought experiment, the directions in which the symmetrical principle can be obeyed are $W \to \infty$, in obedience to the second law in which increasingly many possible arrangements become available globally and $W \to 1$, in which there is a local tendency towards only one possible arrangement as compelled by the continuing local reduction of energy due to the further universal expansion taking place.

It should be recalled and noted that the physical way the increase in $W$ is made manifest is by global proliferation of constituent compartments so that more possible arrangements can take place. This has implication for our local system under experimental observation since we keep the volume fixed. We thus have to accept that what is constant is the volume under observation and not necessarily all the constituent compartments comprising that volume, especially those constituent compartments that are not thermodynamically restricted from participating in the different arrangements available. The implication is that the effect of the global tendency for $W$ to increase is physically felt locally as well, including within the volume under experimental observation. Using the more general description for acceleration-deceleration above, we may reasonably expect to see the two tendencies at play locally, i.e. acceleration towards $W \to \infty$ which holds globally and acceleration towards $W \to 1$, which holds locally. In the same way as we simultaneously describe acceleration-deceleration in the directions of fire and recoil with the gun and bullet, acceleration towards $W \to 1$ represents deceleration to $W \to \infty$ and acceleration towards $W \to \infty$ is deceleration to $W \to 1$. For a system in equilibrium between the two tendencies, we may expect the acceleration and deceleration to each way of obeying the symmetrical principle to occur in turns and this being the case, oscillatory phenomena may be observable. Such observed phenomena showing resistance to collapse into a single arrangement, i.e. $W \to 1$, as dictated by the local thermodynamic necessity would enable the local system "know" that the global system is obeying the symmetrical principle in a different way.

As a kind of simplified summary, wear a thinking cap like Yang's and in the manner of thought experimentalists like Galileo and Einstein shut yourself in a cabin in the early universe (like the local system in Fig.5) equipping yourself with a device capable of monitoring thermodynamic parameters, then make your observations as the universe expands. What can be seen? Energy falls locally, temperature drops and there is cooling. As a result, entropy and disorder also drop locally and some ordering and structural forms are compelled to emerge and condense out of the amorphous environment. All these are not strange and are in keeping with known thermodynamic laws and behavior. How then can the observer "know" whether the global system is behaving in a similar thermodynamic way? The answer as Yang tells us depends on the observation of force. If he does not see force, then he would "know" that the local and global thermodynamic behaviors are the same. If he does see force, he would "know" that the global thermodynamic behavior differs from the local. How then will this force, if seen be expected to look like? The answer Yang may again give us is that since force is a product of two opposing behaviors, there are bound to be some phenomena showing equilibrium between the two opposite thermodynamic behaviors. When we see such phenomena, we would be compelled to wonder what prevents collapse into a single possible way of arrangement in obedience to the local way of keeping the symmetrical principle, since $W \to 1$, when energy and temperature continue reducing towards zero locally as the universe continues expanding? On the other hand, when we see these same phenomena, we would at the same time be puzzled and compelled to ask what prevents arrangement in all the possible ways increasingly made available to the system in obedience to the global way of keeping the symmetrical principle with $W \to \infty$, which we also know as the second law? Our thought experiment tells us that this is what force would look like.

In the real universe, although quantum theory proposes the use of stationary electron waves as a useful solution to the dilemma of atomic orbital stability, there is still some remaining wonder how exactly such stationary waves are formed to keep the electron from collapsing into a single frame of reference with the nucleus. Newton on the other hand could find no satisfactory explanation to prevent his clockwork orbits from spiraling inwards and collapsing under the local influence of the centripetal gravitational force. A centrifugal acceleration was required for stability, to oppose the centripetal acceleration. The questions raised in our thought created universe can therefore also be asked in the real one. What keeps the earth from collapsing into a single frame of reference with the sun? What keeps quarks from collapsing into a single possible arrangement rather than forming a tiny ball and orbiting around each other? On the other hand, we are also compelled to ask what keeps the electron from dancing away and being arranged in all the different possible arrangements available? What prevents the earth and its constituent particles from being arranged in all the available ways in the system? These questions, particularly the aspect why the global tendency towards being arranged in all the possible ways available is not obeyed is explained with the use of force. The other aspect why the local tendency and collapse into a single frame of reference is not obeyed has been more difficult to explain and various ad-hoc mechanisms, such as stationary electron waves, exclusion principles, etc have been employed.

From our thought experiment, we may propose that orbits in the real universe are phenomena manifesting a balance between the local and global thermodynamic tendencies. But for the local tendency, constituents will be arranged in all the possible ways available in the system and we could not know the phenomenon of force or structure. Conversely, without the global tendency for arrangement in more of the different possible ways existing alongside a local tendency towards arrangement in only one possible way, we would not know force as well. An electron in the hydrogen atom would travel the 0.05 nanometre radius and collapse into the nucleus in ~ $10^{-16}$ seconds under the local tendency towards $W \to 1$ alone, while the earth-moon system with an acceleration of 0.0027 ms$^{-2}$ towards each other and ~ 10$^9$ metres apart would collapse into a single frame of reference within 6 days under the influence of the local tendency alone. That such events are forestalled suggests the influence of a global tendency in local affairs. Globally the increase in $W$ manifests as an expanding universe, while locally it manifests as a resistance to orbital collapse.

Parts or constituents of a system possessing a thermodynamic tendency towards $W \to 1$ while obeying the invariance of Boltzmann's constant can be dynamically seen to be tending towards being in the same frame of motion and having only one possible arrangement out of the many possible frames of reference and arrangements. As has been highlighted, this tendency towards being in the same frame of reference is contrary to the dictates of the second law and invariably some type of force is necessarily described. For example, the electron that tends towards being in the same frame of reference and motion as the nucleus makes the description of electromagnetism necessary and in the same way, the earth-based observer who tends to be in the same frame of motion as the earth makes gravitational force necessary to explain the occurrence. In subsequent contemplated papers, we intend to explore this as a thermodynamic basis for explaining experimental observations based on frames of reference, particularly concerning phenomena such as local Lorentz invariance because of the importance such phenomena have assumed in our physics.

This paper is not much concerned with the goal of a unification theory for the fundamental forces, which however remains a topical area of much speculation and study. Nevertheless it is worthy to note that before progress can be made towards unification, we must first know what fundamentally force is and represents before we further examine the different types, their peculiarities and whether one can be convertible to another. There will additionally be the need to examine and resolve how force originated in the real creation experiment, that is, do the forces of nature blend retrospectively into a putative primal force at the beginning, i.e. does a single existing force at the beginning branch out into the observed forces now seen as the universe evolved or was there even no force at all at a beginning from nothing, the variously described forces emerging in obedience to some mathematical or physical principle as the universe evolved?

In our thought experiment, there seemed to be no primal force at the beginning but an invariant symmetrical principle, with structure and the various associated forces seeming to emerge as thermodynamic necessities as the universe evolved. If unification must be pursued, then all the fundamental forces can, in spite of their individual peculiarities, be jointly described as different ways in which the constituents of a system are denied their participation in the increasing number of possible arrangements made available by the second law of thermodynamics because of a need to obey an overriding symmetrical principle in the form of invariance of Boltzmann's constant. This overweening symmetrical principle based on thermodynamics seems to unite and give all the fundamental forces a common reason to exist.

**Conclusion**

The universe is assumed to have had a perfectly symmetrical origin. Support for this belief of maximal order or infinitely low disorder at the beginning can be found in General relativity, the second law of thermodynamics and various cosmological theories particularly those that speculate a beginning from nothing. However, this cherished belief gives rise to the problem of consistently explaining how structure could then be coaxed out of such a smooth and homogenous beginning. In order to provide a basis for the evolution of structure that is now visualized in reality, this cherished scenario has had to be modified on an ad-hoc basis and it has been suggested that for cosmic structure to have evolved, inhomogeneities must have been built right into the beginning without which the universe would have evolved homogenously without the appearance of structure. The essential logic is that for there to be structure today, e.g. in form of galaxies, the structure must have been there from the beginning on a small scale. The two mechanisms usually proposed are that either inhomogeineities must have been present and built into the big bang singularity as part of the initial conditions and it is these that represent the 'seeds' that have now evolved into galaxies [10,11] or that quantum fluctuations occurring during a possible inflationary epoch represent the inhomogeineities that have arisen from the initially homogenous beginning to give rise to cosmic structure [10-13]. The defacto presence of cosmic structure thus seems to necessitate the jettisoning of the cherished idea that the beginning was infinitely homogenous and of perfect symmetry. Aside this is the further puzzle that even if inhomogeineities were built into the beginning, fine-tuning of the universe's expansion would still be required without which structure would not form [11]. This is because structure formation is regarded primarily as a gravitational process rather than a thermodynamic one and the process could be overwhelmed by a very rapid expansion or aborted by a slower expansion rate.

From the results of our thought experiment, we see that based on symmetry considerations and the thermodynamic principles at play, the appearance and evolution of force and structure can qualitatively be described and would be inevitable during the universe's transformation even if we still retain our cherished belief that the universe had a perfectly symmetrical and homogenous beginning.

The universe created from our thought experiment seems to substantially share many characteristics with the real universe and it would be tempting to state that indeed the same physical processes apply in both, especially as both are governed by the same thermodynamic laws, equations and physical principles.

**References**

1. Ojo, A., Creating a universe from thermodynamic thought experiments I, arxiv:physics/0607017.

2. Perrot, P., *A to Z of Thermodynamics*. Oxford University Press, London, 1998

3. Fermi, E., *Thermodynamics*, Prentice-Hall, Englewood-Cliffs, 1937

4. Callen, H.B., *Thermodynamics and an Introduction to Thermostatistics*, Wiley, New York, 1985

5. Clausius, R., Uber verschiedene fur die Anwendung bequeme Formen der Hauptgleichungen der mechanischen Warmetheorie. *Annalen der Physik und Chemie*, 125:353–400, 1865.

6. Boltzmann, L., Weitere Studien uber das Warmegleichgewicht unter Gas-Molekulen. *Sitzungsbericht der Akadamie der Wissenschaften, Wien*, 66:275-370, 1872.

7. Newton, I., *Principia*, 1687

8. Yang, C.N., *Elementary Particles*, Princeton University Press, Princeton, New Jersey, 1961

9. Ferris, T., *Coming of Age in the Milky Way*, William Morrow, New York, 1988.

10. Rees, M.J., The emergence of structure in the universe: galaxy formation and dark matter. In *300 Years of Gravitation*, Eds. Hawking, S.W. and Israel, W., Cambridge University Press, Cambridge, 1987.

11. Rees, M.J., *Just Six Numbers*, Weidenfeld & Nicolson, London, 1999.

12. Blau, S.K. and Guth, A.H., Inflationary cosmology. In *300 Years of Gravitation*, Eds. Hawking, S.W. and Israel, W., Cambridge University Press, Cambridge, 1987.

13. Linde, A., Inflation and quantum cosmology. In *300 Years of Gravitation*, Eds. Hawking, S.W. and Israel, W., Cambridge University Press, Cambridge, 1987.